\documentclass[twocolumn,final]{svjour3} 
\smartqed  

\usepackage[export]{adjustbox}

\usepackage{multirow} 
\usepackage{rotating}
\usepackage{kantlipsum} 
\usepackage{commath}
\allowdisplaybreaks
\usepackage{mathtools}  
\usepackage{algpseudocode}
\usepackage{algorithm}
\usepackage{verbatim}
\usepackage{xr-hyper} 

\usepackage{amsmath} 
\usepackage{epstopdf}
\usepackage{wrapfig}
\usepackage{latexsym}
\usepackage{amssymb}
\usepackage{amsfonts}

\usepackage{graphicx}
\usepackage{latexsym}
\usepackage{booktabs}
\usepackage{subcaption} 
\usepackage{breqn}
\algnewcommand\algorithmicinput{\textbf{INPUT:}}
\algnewcommand\INPUT{\item[\algorithmicinput]}
\algnewcommand\algorithmicoutput{\textbf{OUTPUT:}}
\algnewcommand\OUTPUT{\item[\algorithmicoutput]}
\usepackage[table]{xcolor}
\usepackage{rotating} 
\usepackage[numbers]{natbib}

\begin{document}

\title{CECT: Computationally Efficient Congestion-avoidance and Traffic Engineering in Software-defined Cloud Data Centers}

\titlerunning{CECT: Computationally Efficient Congestion-avoidance and Traffic Engineering}
	\author{M.M.~Tajiki,\and B.~Akbari,\and M.~Shojafar,\and S.H.~Ghasemi,\and M.L.~Barazandeh,\and N.~Mokari,\and L.~Chiaraviglio,\and M.~Zink}

\institute
    {
        MM. Tajiki, S. Hesamoddin Ghasemi, M. Latifi Barazandeh, N. Mokari, B. Akbari \at lectrical and Computer Engineering,\\ Tarbiat Modares University, Tehran, Iran\\\email{\{mahdi.tajiki, h.qasemi, mahdi.barazandeh, b.akbari, mokari\}@modares.ac.ir}
	\and
		M. Shojafar, L. Chiaraviglio  \at CNIT, Department of Electronic Engineering,\\ Tor Vergata University of Rome, Rome, Italy.\\\email{mohammad.shojafar@cnit.it, luca.chiaraviglio@uniroma2.it}
	\and 
	    Michael Zink \at Electrical and Computer Engineering,\\University of Massachusetts Amherst, Amherst, USA\\\email{mzink@cas.umass.edu}
	}
\date{Received: 3 July 2017 / Revised: 13 November 2017/ Accepted: XX XX}
\maketitle
\begin{abstract}
The proliferation of cloud data center applications and network function virtualization (NFV) boosts dynamic and QoS dependent traffic into the data centers’ network. Currently, lots of network routing protocols are requirement agnostic, while other QoS-aware protocols are computationally complex and inefficient for small flows. In this paper, a computationally efficient congestion avoidance scheme, called \textit{CECT}, for software-defined cloud data centers is proposed. The proposed algorithm, CECT, not only minimizes network congestion but also reallocates the resources based on the flow requirements. To this end, we use a routing architecture to reconfigure the network resources triggered by two events: 1) the elapsing of a predefined time interval, or, 2) the occurrence of congestion. Moreover, a forwarding table entries compression technique is used to reduce the computational complexity of CECT. In this way, we mathematically formulate an optimization problem and define a genetic algorithm to solve the proposed optimization problem. 
We test the proposed algorithm on real-world network traffic. Our results show that CECT is computationally fast and the solution is feasible in all cases. In order to evaluate our algorithm in term of throughput, CECT is compared with ECMP (where the shortest path algorithm is used as the cost function). Simulation results confirm that the throughput obtained by running CECT is improved up to 3x compared to ECMP while packet loss is decreased up to 2x. 
\end{abstract}


\section{Introduction}\label{sez:1}

Network Function Virtualization (NFV) has drawn significant attention from industry, government, and acad-emia to improve flexibility and reduce the time to market of new services. Some of these services have a chain of functions (e.g., firewall and load balancer) which need the network to guarantee the required Quality of Service (QoS) constraints. On the other hand, the data centers used to create cloud services represent a significant investment in capital outlay and ongoing costs. Therefore, the cloud data center services are highly adapted at the present time. The dynamic nature of the cloud data center traffic (e.g., VM motion) necessitate support for the diverse class of QoS requirements. Not surprisingly, these QoS requirements have to be guaranteed by the network routing protocol. Additionally, the enormous and dynamic network traffic which is communicating via the network infrastructure imposes congestion in the network links. Clearly, this effect has to be dynamically addressed by the routing protocols in Software-defined Cloud Data Centers (SCDC). 
In a sequel, the main aim of this paper is to dynamically and efficiently reallocate resources in a way that i) guarantees QoS requirements of different applications; and; ii) protectively prevent congestion and resource waste.  

In this context, several questions arise, like: Is it possible to propose an approach that considers the impact of flows routing among each other? How to model an SCDC and services to evaluate the flow requirements? How to implement a real-time computationally efficient method to preserve SCDC QoSs? The answer to these questions is the goal of the paper. 

More in detail, we introduce a dynamic and computationally efficient resource reallocation scheme called Computationally Efficient Congestion avoidance and Tr-affic Engineering (CECT) in which we guarantee the minimum bandwidth for a specific flow. The main contributions are as follows:
\begin{itemize}
	\item[i)] The proposed scheme not only maximizes the network throughput but also guarantees the requested QoS level. Since the traffic flow requirements change over time, the mentioned scheme dynamically reallocates the resources in a predefined time period. In order to solve the corresponding optimization problem, two schemes are proposed. The \textit{first one} maximizes the total network throughput, where its computational complexity is high. The \textit{second one} is a low computational complexity meta-heuristic method that finds a near-optimal solution.
	\item[ii)] To overcome the resource fragmentation in networks with big flows, we consider the impact of each flow on other flows, i.e., in the corresponding optimization problem the rerouting of all flows must be performed simultaneously. Additionally, in order to improve the congestion avoidance as well as increasing the network throughput, multi-path routing is supported in the proposed schemes. Hence, different flows from a similar source to a similar destination can be rerouted via various paths.
	\item[iii)] To make a tradeoff between computational complexity and performance, the granularity of network rescheduling is adjustable. To this end, we introduce a flow table entry compression technique that makes a tradeoff between the optimality gap and the computational complexity of the solution. In our design, the granularity can be the exchanged information of  ``a special application in a server with another application on a different server'' or ``all communications from one data center to another one''.
	\item[iv)] We implement CECT in the MiniNet emulator \cite{29mininet}, by considering a realistic network traffic and a realistic fat-tree network topology. Besides, in order to evaluate the impact of flow size on the performance of CECT, we implement a packet generator to generate traffic patterns with micro, small, medium, and big flows. 
	
\end{itemize}

The full evaluation of other QoSs features (such as the queuing delay, delay variation (jitter), quality of user experiments in SCDC, and the mapping of such parameters in other types of the networks such as WAN) will be some interesting branches of future research.

The rest of the paper is organized as follows. Section \ref{sez:2} presents the most recent existing literature. In Section \ref{sez:3}, we present the main functional blocks of the proposed CECT architecture for SCDC. Afterward, in Section \ref{sez:4}, we detail the problem formulation, while Section \ref{sez:5} proposed the bio-inspired scalable solution. After detailing the tested application scenarios and performance metrics in Section \ref{sez:6}, the performance of CECT algorithm is presented and compared with the corresponding one of the ECMP algorithm \cite{23-5454063} in Section \ref{sez:7}. Finally, in Section \ref{sez:8}, we summarize the main attained results and give some hints for future research.

\section{Related work}\label{sez:2}
In the following, we will briefly discuss the main literature engaged in SCDC.
\subsection{Congestion Avoidance/Control Methods}\label{sez:2.1}
In the literature, different works on congestion avoidance/control and traffic engineering have been presented in the past. {Authors in \cite{23-5454063} presented a multi-path routing technique, ECMP, to perform static load splitting among flows across 8 to 16 multi paths. It is required to deliver high bisection bandwidth for larger data centers. ECMP is applied in current switches in which that are tuned and configured with several possible forwarding paths for a given subnet. In other words, when a packet with multiple candidate paths arrives, it is forwarded on the one that corresponds to a set of selected fields of that packet's headers and modules the number of paths. ECMP does not account for flow bandwidth in making allocation decisions, which can lead to oversubscription that CECT matters this issue.} 

In \cite{1do2016sliding}, a congestion control scheme classifies the network traffic into two classes of \textit{ordinary} and \textit{premium} flows. More in depth, authors consider a non-linear network model based on the fluid flow theory that is able to cope with both the physical network resource constraints and unknown time delays associated with networking systems. However, the proposed scheme does not embed any traffic engineering scheme, therefore, {it does not specify routes for the network flows that the proposed CECT method does.} The authors of \cite{2su2016optimization} proposed a traffic engineering approach for networks in which the link capacity and class of service requirements may vary with time. Their scheme does not route flows, however, it produces some control laws which can be used for routing. In \cite{3otoshi2015traffic}, a traffic prediction algorithm is exploited to prevent network congestion before it happens. Still, the approach is not applicable in networks with unpredictable traffic pattern. {Instead, CECT schedules the resources and flows according to the complete view of the system state and is unbounded to any traffic patterns.}
	
The authors of \cite{4lu2015sdn} propose an SDN-based TCP congestion control mechanism at the client side. They focus on long-lived flows and reduce the sending rate by adjusting the TCP receive window of ACK packet after OpenFlow switch triggered a congestion message to the controller. Similarly, authors in \cite{5gholami2015congestion} proposed a method to control the congestion in SCDCs based on the OpenFlow protocol. Their method monitors the port statistics of the OpenFlow-enabled switches and reroutes some flows in the congested links. Both \cite{4lu2015sdn} and \cite{5gholami2015congestion} do not consider QoS requirements of different flows. In other words, they assume that all flows have similar requirements. {Instead CECT considers
the flows features and SDN resources properties dynamically.} 

In \cite{6tajiki2017optimal}, a QoS-aware resource allocation algorithm which guarantees a minimum overhead on the network during reprogramming phase. The authors mathematically formulate the optimization problem of flow routing in the data center networks and solve it using binary linear programming. The most challenging part of their method is the high computational complexity of solving the mentioned problem which makes the proposed algorithm versy inefficient in medium and large scale data centers. Moreover, the authors of \cite{7tajiki2016qrtp}, propose a routing algorithm for SCDCs based on traffic prediction. They mathematically formulate the routing problem and propose two schemes to solve it (an exact solution which has a high computational complexity and a suboptimal but fast one). {This paper presents a flavor of the significant results and ongoing work, but it is applicable only in predictable networks whilst CECT is not only applicable in the predicting network, but also it is used in unpredicted and any shape of the networks (i.e., based on proof of concept presented in the simulation results). }

In addition, authors in \cite{8shetty2016optimizing} present a network-aware resource reallocation technique, in which they use the network topology characteristics of the data center to minimize the maximum latency in communication between VMs. They incorporate the resource heterogeneity by including the computational and communication requirements in the proposed technique. {The main focus of the work is on heterogeneity of computational requirements for VMs in CDC, and did not consider heterogeneity of the network bandwidth and the other computational requirements for VMs in CDC. Instead, the proposed method, CECT, covers these limitations and provides load balancing routing to the incoming flows.} 

In addition, the authors of \cite{9zhang2016iterative} adopt a two-phase flow embedding approach with an iterative traffic engineering algorithm to address the resource reallocation problem lying in the multimedia communication systems. Some other works such as \cite{10mushtaq2015qoe} focus on providing QoS for voice over IP (VoIP) traffics and simultaneously optimizing the power efficiency. {In CECT, we not only cover the VoIP traffic class but also consider several types of traffic, e.g., FTP, high definition (HD) video stream that can be applied in 5G network.}
\vspace{-10px}
\subsection{QoS-aware Routing Methods (Single Class of Traffic)}\label{sez:2.2}
Different works that focus on multimedia and use flow rerouting to guarantee the QoS parameters have been presented in \cite{11egilmez2013optimization,12egilmez2012openqos,13egilmez2012distributed} and \cite{14egilmez2011scalable}. In detail, the authors of \cite{11egilmez2013optimization} formulate the dynamic QoS routing problem as a Constrained Shortest path (CSP) problem. In this way, they represent the entire network as a simple graph and define a cost function based on the QoS parameters. The proposed solution improves the QoS of video streaming. However, it cannot support different classes of QoS that is covered in CECT. In \cite{12egilmez2012openqos}, the authors form in a group the incoming flows as multimedia and data flow, where the multimedia flows are routed via QoS guaranteed paths. However, the data flow remains on traditional shortest-paths. {Instead in CECT, we cover various types of incoming traffic.} {Besides, authors in \cite{shojafar2016adaptive} address multimedia data processing with computationally intensive tasks and exchange of a big volume of data flow via QoS guaranteed paths per time period and introduce a general framework called \textit{MMGreen} to ensures QoSs of the user flows and achieves maximum energy saving and attains green cloud computing goals in a fully distributed fashion by utilizing the DVFS-based CPU frequencies. Although MMGreen is novel and interesting, but compared to CECT in routing scheme, it does not cover some QoSs such as the utilization of the links between the SDN switches.}

Moreover, the authors of \cite{13egilmez2012distributed} propose a distributed QoS routing architecture for video streaming. They use the OpenFlow features to implement their scheme in a multi-domain environment. Finally, \cite{15civanlar2010qos} devises a simple analytic framework and an experimental platform to transfer the video streaming. In their framework the video stream has a base layer, which is modeled as a QoS flow, and multiple enhancement layers, which are treated as best-effort flows.

\subsection{QoS-aware Routing Methods (Multiple Class of Traffic)}\label{sez:2.3}
The main challenging deficiency of the mentioned schem-es is not to support different classes of QoS, i.e., these solutions only focus onmultimedia flows and ignore other types of data flows. In contrast, there are many works considering different types of data flows such as \cite{16kulkarni2012new,17liang2015one,18leela2011multi,19ghosh2013scalable,20wang2014sdn,21ongaro2014enhancing} and \cite{22zhao2015qos}. Particularly, authors in \cite{16kulkarni2012new} propose a new QoS routing algorithm for Multi-Protocol Label Switching (MPLS) networks. The paths are selected based on critical links so as to minimize interference with the future requests. {However, the solution \cite{16kulkarni2012new} is designed for MPLS and it can not be used for other types of network.} In addition, the main focus of \cite{18leela2011multi} is to route the flows with the QoS constraint using genetic algorithms. In this way, authors propose a heuristic for unicast routing to find feasible path satisfying the flows requirements. In the same way, \cite{17liang2015one} guarantees the QoS by defining a new measure called \textit{path weight} which is minimized by the aid of ant colony system. Both \cite{18leela2011multi} and \cite{17liang2015one} route each flows separately. {These methods have three set of limitations. First, they do not guarantee the efficiency of the selected path due to the minimum requirement constraint applied for termination conditions. Second, they do not guarantee the end-to-end performance. In other words, the proposed routing algorithms do not exploit the capability of routing all flows simultaneously, i.e., it is impossible to reroute a flow considering the possible routes of other flows. Third, they examined single flow performance, and unable to handle multiple flows with different QoS requirements. Consequently, these schemes are not applicable for a comprehensive network reconfiguration.} 

In addition, the scheme reported in \cite{19ghosh2013scalable} explores scalable architectures that jointly optimize rate control and routing. Since the goal of this work is to perform rate control, the proposed approach distributes information and computation across multiple tiers of an optimization machinery. Similarly, in \cite{20wang2014sdn}, the authors present a resource allocation scheme for inter-data center communication with multiple traffic classes. {Although \cite{19ghosh2013scalable} and \cite{20wang2014sdn} are practical for inter-data center communications, they are impractical for intra-AS (i.e., inter-autonomous system) network resource allocation, e.g., intra-cloud data centers.}
\begin{figure*}[!htbp]
	\includegraphics[width=\textwidth]{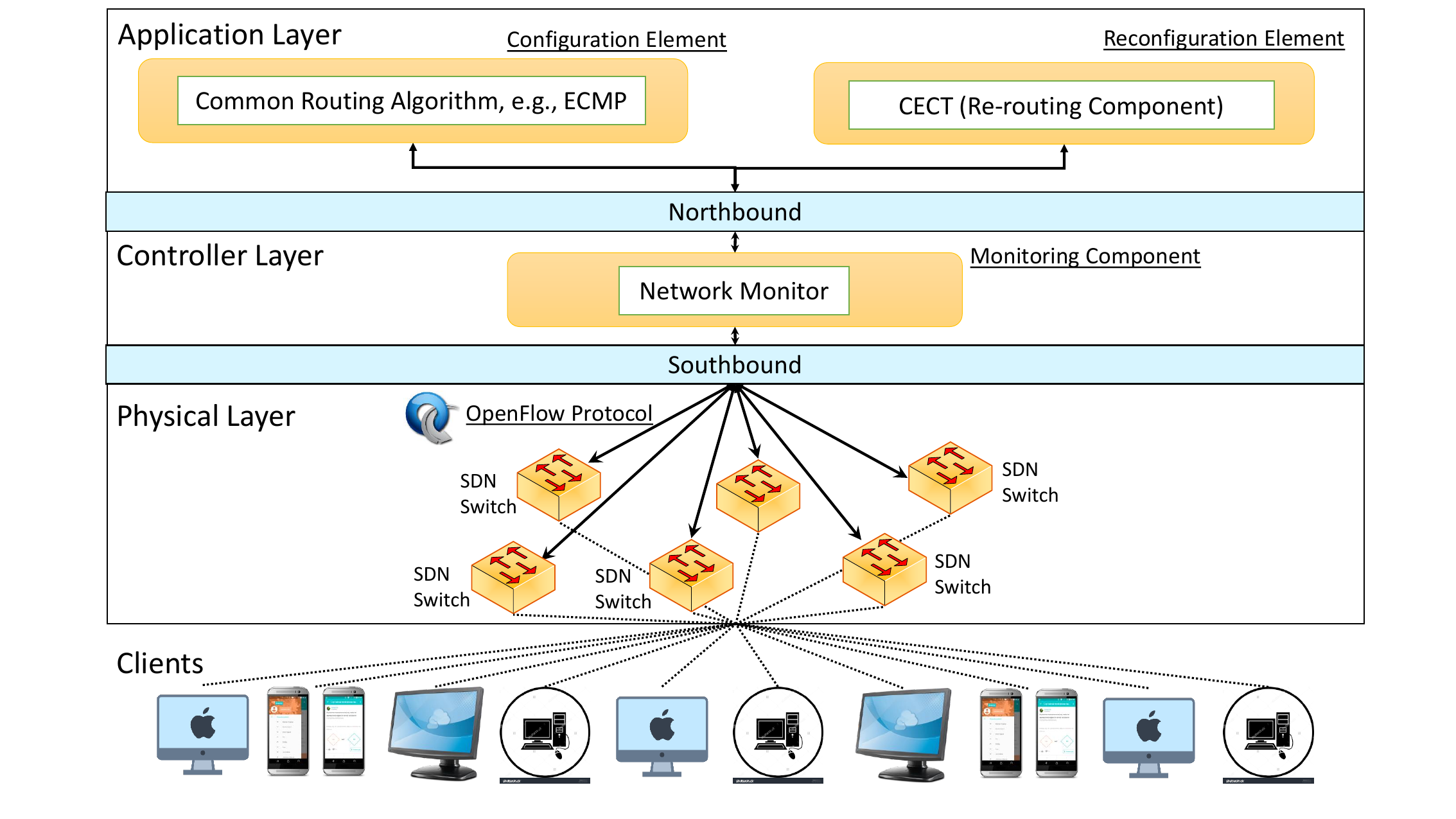}
	\caption{The proposed architecture.}
	\label{fig:fig1}
\end{figure*}
The paper \cite{21ongaro2014enhancing} deploys the SDN features to manage the differentiating network services with QoS satisfaction. Their problem formulation is in form of integer linear programming (ILP). The weak point of this scheme is its computational complexity which makes it impractical for medium and large scale networks. In \cite{22zhao2015qos}, a QoS-aware routing mechanism is proposed to balance the network load of industrial Ethernet. In particular, authors exploit the ant colony method to obtain a path for data transmission with different QoS requirements. Since the meta-heuristic approaches may return unfeasible solutions, they need to be discussed from the validity perspective of the solution. However, the time complexity and the validity of their scheme are not discussed.

\section{Reference Architecture}\label{sez:3}
In Fig. \ref{fig:fig1}, we describe the considered architecture. In particular, we assume a software-defined cloud network where a logically centralized controller coordinates the network. The switches are all OpenFlow-enabled and the protocol used for communication between the swit-ches and the controller is OpenFlow. Therefore, the controller may query the switches for network topology and current traffic matrix. On the other hand, there are $k$ different classes of traffic flows with different QoS requirement. In addition, the flows are highly dynamic and there are some big flows in the network. In the case of arrival of a new flow, a conventional routing scheme like ECMP \cite{23-5454063} is applied. In order to decrease network congestion, the controller reconfigures the network, i.e., some flows are rerouted. To this end, the resources are reallocated based on some predefined measures such as time periods or the event of high packet loss in the network. Table \ref{table1} reports the main notations of the paper.

In order to maximize the total network throughput as well as guarantee the QoS requirement of flows, it is necessary to dynamically reallocate the resources with the dynamic pattern of network traffic. In this way, we design a mathematical model that considers the network topology, the flow requirement matrix, and the flow specifications as input, and finds a routing matrix satisfying the QoS constraint and minimizing network congestion. The network topology is given by the matrix $B_{N_L\times N_L}$ where $B_{(i,j)}$ determines the bandwidth of the link from the switch $i$ to the switch $j$. The number of flows and OpenFlow-enabled switches is $N_F$ and $N_L$, respectively. The routing matrix $A_{N_L\times N_L\times N_F}$ specifies the path selected for each flow, e.g., if $A_{(i,j)}^f \in \{0,1\}$ is equal to 1 then the flow $f$ crosses the link or $i\rightarrow j$. The flow requirement matrix $C_{1\times N_F}$ specifies flows guaranteed requirements based on the corresponding class. The $i$-{th} row of the flow requirement matrix defines the guaranteed bandwidth for each flow. 
\begin{table}[!htbp]
	\begin{center}
		\caption{Main Notation.}
		\begin{tabular}{|c|l|}
			\hline
			\textbf{Symbol}& \textbf{Definition} \\
			\hline\hline
			\multicolumn{2}{|c|}{\textbf{Mathematical Parameters}}\\\hline\hline
			$N_L$& Number of switches\\\hline
			$N_F$& Number of flows\\\hline
			$B$& $N_L \times N_L$ matrix denoting the links bandwidth\\\hline
			$R$& $1\times N_F$ vector denoting flows requirement \\\hline
			$s$& $1$$\times$$N_F$ vector denoting source switch of flows\\\hline
			$d$& $1$$\times$$N_F$ vector of destination switch of flows\\\hline\hline
			\multicolumn{2}{|c|}{\textbf{Metaheuristic Parameters}}\\\hline\hline
			$PO$& Population as set of solutions\\\hline
			$CH$& Chromosome set of x-paths\\\hline
			$XP$& A gene which is a x-path\\\hline
			$V$& A set of switches (nodes)\\\hline
			$E$& A set of edges (links)\\\hline
			$R$& A set of all x-paths\\\hline\hline
			\multicolumn{2}{|c|}{\textbf{Decision Variable}}\\\hline\hline
			$A$& $N_L$ $\times$ $N_L$ $\times$ $N_F$ routing matrix\\ \hline
			$\mu$& Maximum link utilization\\  		
			\hline
		\end{tabular}
		\label{table1}
	\end{center}
\end{table}

It should be mentioned that CECT is a secondary routing algorithm, which means that there is a primary routing algorithm along with it. In other words, in order to minimize the routing delay of new arrival flows, CECT uses a conventional routing algorithm (as an example, ECMP) to route the flows separately. Thereafter, if the link utilization of some parts of the network exceeds a predefined threshold, CECT algorithm is invoked and some flows are rerouted to prevent network congestion. {The average links utilization in different types of networks are different, in \cite{benson2012new}, several data center traffics are investigated and the flows characteristics are well studied. Based on their study, links with 70 percent and higher utilization are considered as hot-spot links. We followed the same setting. It should be mentioned that using high values as the threshold makes the algorithm more quick while it increases the probability of congestion for burst traffic. On the other hand, considering a low value balances the load across the network while it increases the execution time.} As it can be seen in Fig. \ref{fig:fig1}, in our architecture the routing of new arrival flows is done using existing routing algorithms while CECT is used to reroute flows for traffic engineering purposes and minimizing network congestion. 
\section{Problem Formulation}\label{sez:4}

The main objectives of this paper are to efficiently and dynamically reallocate resources in a way that i) the QoS requirements of different applications are guaranteed, ii) the resource waste and congestion are proactively prevented, and, iii) the computational complexity of rescheduling process is minimized. As a consequence, the routing matrix should be calculated in a way that the mentioned constraints are satisfied. To this end, the routing matrix $A_{N_L\times N_L\times N_F}$ and $\mu$ can be obtained such that the network rescheduling overhead is minimized subject to the QoS constraints and the flow conservation constraints.  In the following, we present the formulation of the considered problem.

\subsection{Capacity Constraint}\label{sez:3.1}
The link load is guaranteed to be smaller than the maximum target utilization $\mu$ by the following constraint:
\begin{equation}\label{eq:eq1}
\sum_{f=1}^{N_F} A_{(i,j)}^f R_f \leq \mu B_{(i,j)},\quad \forall i,j \in \{1,\ldots,N_L\},
\end{equation}

Specifically, the l.h.s of \eqref{eq:eq1} calculates the sum of the guaranteed bandwidth of all flows crossing a specific link. The right-hand side specifies the maximum predefined allowable link bandwidth.

\subsection{Source and Destination Constraints }\label{sez:3.2}

The flows are prevented from returning to the source switches via Equation \eqref{eq:eq2}. As mentioned earlier, $A_{(i,s_f)}^f$ is one if and only if the flow $f$ crosses the link that connects switch $i$ to the source switch of $f$ called $s_f$. We then impose the following constraints:

\begin{equation}\label{eq:eq2}
\sum_{i=1}^{N_L} A_{(i,s_f)}^f =0,\quad \forall f \in \{1,\ldots,N_F\},
\end{equation}
\begin{equation}\label{eq:eq3}
\sum_{i=1}^{N_L} A_{(d_f,i)}^f =0,\quad \forall f \in \{1,\ldots,N_F\},
\end{equation}
For each flow, Equation \eqref{eq:eq2} forces the summation of $A_{(i,s_f)}^f$ (for all $i$) to be zero. In other words, none of the flows can cross the link between any switch to the source switch of that flow. On the other hand, \eqref{eq:eq3} makes the flows to stay on the destination switches. The aforementioned constraints prevent the flows from entering an invalid switch. Moreover, equations \eqref{eq:eq4} and \eqref{eq:eq5} prevent flows from staying in an invalid switch. These equations force the flows to leave the origin switches and enter to the destination switches, respectively.
\begin{equation}\label{eq:eq4}
\sum_{i=1}^{N_L} A_{(s_f,i)}^f =1,\quad \forall f \in \{1,\ldots,N_F\},
\end{equation}
\begin{equation}\label{eq:eq5}
\sum_{i=1}^{N_L} A_{(i,d_f)}^f =1,\quad \forall f \in \{1,\ldots,N_F\},
\end{equation}
Equation \eqref{eq:eq4} guarantees the flows to cross from exactly one of the source switch outgoing link. Similarly, equation \eqref{eq:eq5} is considered for the incoming links of destination switch.

\subsection{Flow Conservation and Loop Prevention Constraints}\label{sez:3.3}
If a switch is neither source nor destination of a flow, the flow must leave that switch after it moves in. This restriction is applied by Constraint \eqref{eq:eq6} via balancing the amount of traffic entered to the switch with the amount of traffic left it.
\begin{equation}\label{eq:eq6}
\begin{split}
\sum_{i=1}^{N_L} A_{(i,j)}^f =\sum_{i=1}^{N_L} A_{(j,i)}^f,\quad \forall f \in \{1,\ldots,N_F\},\\ \forall i \in \{1,\ldots,N_L\}-\{s_f, d_f\},
\end{split}
\end{equation}
\begin{equation}\label{eq:eq7}
\sum_{i=1}^{N_L} A_{(i,j)}^f \leq 1,\: \forall f \in \{1,\ldots,N_F\},\:\forall i \in \{1,\ldots,N_L\},
\end{equation}
\begin{equation}\label{eq:eq8}
A_{(i,j)}^f \in \{0,1\},\: \forall f \in \{1,\ldots,N_F\},\:\forall i,j \in \{1,\ldots,N_L\},
\end{equation}
Constraint \eqref{eq:eq7} assures there is no loop in the new routing matrix. It prevents flows from returning to a switch that is met in the past. Finally, we express $A^f_{(i,j)}$ as a binary variable.

\subsection{Objective Function}\label{sez:3.4}
In order to minimize network congestion in case of the existence of burst traffic, the objective function is minimizing the maximum link utilization $\mu$. More formally, we have:
\allowdisplaybreaks
\begin{align}
&{\qquad \qquad    \min \quad\mu},\label{eq:eq88}\\	
&\qquad \qquad\text{Subject to:}\nonumber \\
&\text{\qquad \qquad Constraints \eqref{eq:eq1}-\eqref{eq:eq8}}\label{eq:eq881}.
\end{align}
Indeed, we are interested to find the routing matrix $A$ while minimizing $\mu$. 
The problem belongs to the class of Mixed Integer Linear Programming (MILP) formulations, and it is NP-Hard \cite{24guck2016function}. Therefore, we rely on a heuristic approach which is detailed in the next section.  
\section{CECT Algorithm}\label{sez:5}
{In this subsection, the meta-heuristic method called CECT (based on the genetic algorithm \cite{eiben1994genetic}) is precisely described and represented as an algorithm in Alg.~\ref{CECT}. In brief, CECT pre-computes some feasible paths for each flow and assigns a random path to each flow. In this way, a collection of different solutions are constructed and each solution is ranked based on the constraints mentioned in the previous section. Then, CECT uses the roulette wheel algorithm to select some solutions as the ancestors of the next generation. The new population is generated by applying uniform crossover and multipoint mutation on these ancestors. This process is applied to each generation till a solution violating no constraint is found or a predefined threshold for the number of iterations is met. 
}

\begin{figure}
	\centering
	\begin{subfigure}[t]{\columnwidth}
		\centering
		\includegraphics[width=.5\columnwidth]{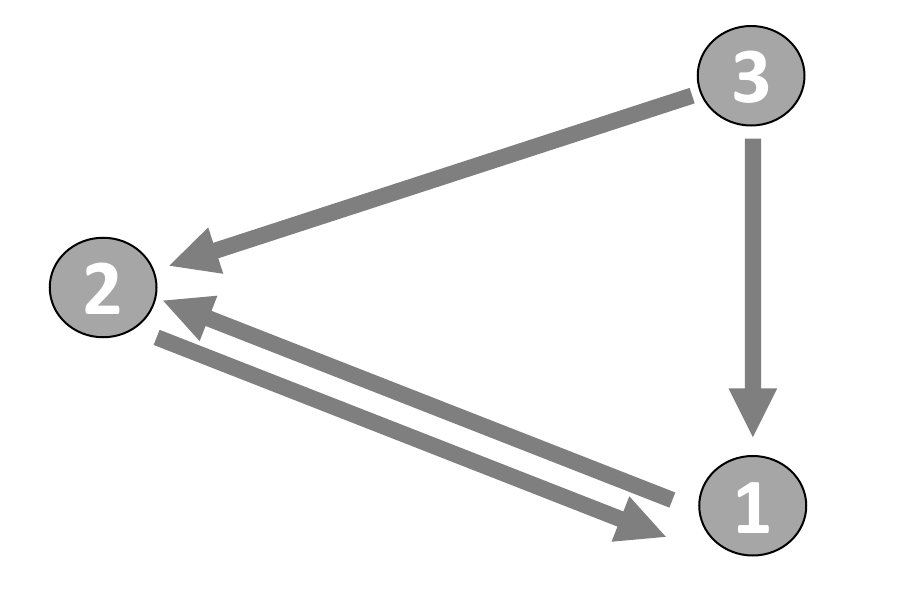}
		\caption{Sample topology with 3 nodes.}
		\label{fig:fig2a}
	\end{subfigure}
	\hfill
	\begin{subfigure}[t]{\columnwidth}
		\centering
		\includegraphics[width=.5\columnwidth]{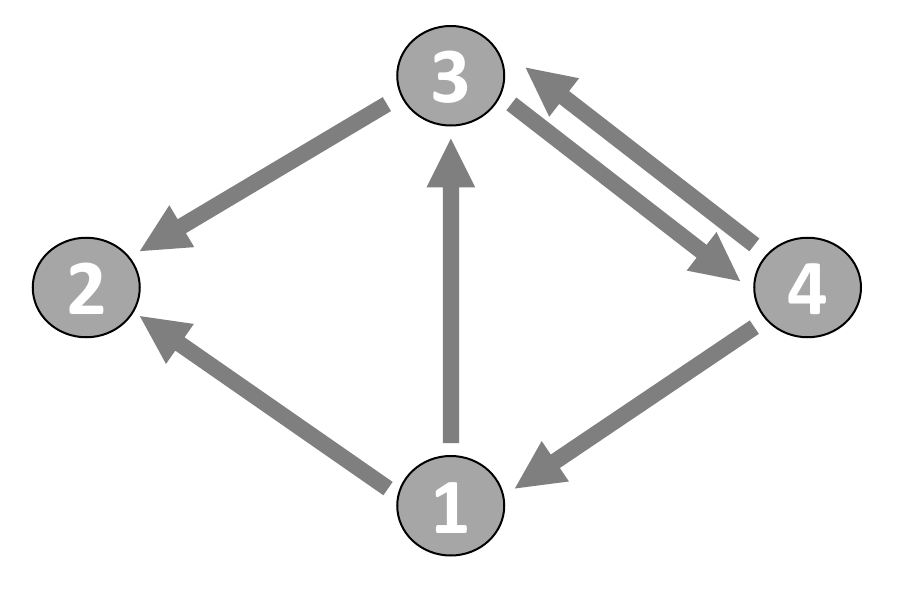}
		\caption{Sample topology with 4 nodes.}
		\label{fig:fig2b}
	\end{subfigure}
	\caption{Topology samples.}
	\label{fig:fig2}
\end{figure}

{In Alg.~\ref{CECT}, the first line computes all x-paths (all paths with a length lower than x) in an offline manner (more details will be provided in Alg.~\ref{PCXP} and Section \ref{sez:4.1}). The second line randomly assigns an x-path to each flow based on the source and destination of flows and x-paths. This means that an x-path which has $(a,b)$ as its (source, destination) cannot be assigned to a flow where the (source, destination) is $(c,d)$ if $(c,d)\neq (a,b)$. Line 4 makes a loop until finding a solution which satisfies all constraints. In other words, the algorithm seeks for a solution with no congestion. It should be mentioned that in two cases the loop breaks after some predefined iterations: i) when the requests are more than the resources (i.e., there is no a solution ensuring all constraints), and, ii) when the proposed algorithm can not find the optimal solution (i.e., in order to prevent an infinite loop). In the next step,  we find the fitness function (i.e., using function \textit{FF} which is precisely described in Section \ref{sec:SelectionStructure}) for each solution (which is called a chromosome, Section \ref{sec:chrom}) for current generation (line 5 of the algorithm). Then, some chromosomes are selected using the roulette wheel algorithm (Section \ref{sec:SelectionStructure} to produce the next generation (line 6). Note that, the number of the selected chromosomes is equal to the number of population. Line 8 protects the best chromosome (best fitness value) from further changes (which is known as elitism in the context of genetic algorithms). In line 9, the selected chromosomes are sent to the uniform crossover function (Section \ref{sec:cross}) to produce the next generation. During this step, the parents are replaced with their children.}
\begin{algorithm}
	\caption{CECT Algorithm}
	\label{CECT}
	\allowdisplaybreaks
	\begin{algorithmic}[1]
		\break\INPUT {$G=<V,E>$, \textit{threshold}}
		\OUTPUT {$A_{(i,j)}^f$, $\forall\: i,j$ }
		\State{\textit{Precompute} all x-paths}
		\State{Randomly select feasible path for each flow (paths with similar source and destination with the flow)}
		\State{$itr=0$;}
		\While{not(Eqs. \eqref{eq:eq1}-\eqref{eq:eq7}) \&\& ($itr$ $<=$ \textit{threshold})}
		\State{\textit{FF($CH_i$)} for each $CH_i\in$ \textit{PO}}
		\State{$S=$\textit{Roulette\_wheel\_selection(PO)};}
		\For{$i=1$;$i<=size(S)$; $i+=2$}
		\If{$CH_i \neq best(FF(S))$} 
		\State{\textit{Uniform\_Crossover}($CH_i$,$\:CH_{i+1}$);}
		\State{\textit{Multipoint\_mutation}($CH_i$);}
		\State{\textit{Multipoint\_mutation}($CH_{i+1}$);}
		\If{no improvement in the solutions for $k$ iterations}
		\State{$mut=mut_{max}$;}
		\Else
		\State{$mut=mut_{min}$;}
		\EndIf
		\EndIf
		\EndFor
		\EndWhile\\
		\Return{$A_{ij}^f$}
	\end{algorithmic}
\end{algorithm}

{Line 10 and 11 are supposed to mutate the newly generated population (Section \ref{sec:mut}). It should be mentioned that two reasons may stop the enhancement of the best chromosome: i) falling into a local optimum, and ii) finding the optimal solution. Therefore, in line 13 of the algorithm the mutation rate is increased to $mut_{max}$ if no improvement is seen in the best chromosome after $k$ iterations. We do this because if the population is in a local optimum then increasing the mutation rate helps the algorithm to escape the local optimum. If the the algorithm is not in a local optimum, in line 15 of the algorithm, the mutation rate is returned to $mut_{min}$ to find an optimum. On the other hand, finding the global optimal solution ends the algorithm (line 20 of the algorithm).}

In the following, we first detail a simple case study for solving CECT problem. Then, we detail the different features of CECT which are the structure of the chromosomes and the subroutines of selection, crossover, and mutation.

\subsection{Preliminaries}\label{sez:4.1}
Consider an x-path is a path with $y$ hops $y\leq x$, e.g., in Fig. \ref{fig:fig2} the set of 3-path is \{\{1$\rightarrow$2\}, \{2$\rightarrow$1\}, \{3$\rightarrow$1\}, \{3$\rightarrow$2\}, \{3$\rightarrow$2$\rightarrow$1\}, \{3$\rightarrow$1$\rightarrow$2\}\}. CECT pre-computes all x-paths (for a predefined $x$) and marks each path as an unique number starting from 1, e.g., \{1$\rightarrow$2\} is marked as 1, \{2$\rightarrow$1\} as 2, and so on. 
Tables \ref{table2} and \ref{table3} contain all entries in the set of 3-path for the sample topologies (i.e., see Fig. \ref{fig:fig2a} for Table \ref{table2} and Fig. \ref{fig:fig2b} for Table \ref{table3}, respectively) along with their labels.

\begin{table}
	\begin{center}
		\caption{3-paths for sample topology depicted in Fig. \ref{fig:fig2a}.}
		\begin{tabular}{|c|c|}
			\hline
			\textbf{Label}& \textbf{Path} \\
			\hline
			1&\{1$\rightarrow$ 2\}\\
			2&\{2$\rightarrow$ 1\}\\
			3&\{3$\rightarrow$ 1\}\\
			4&\{3$\rightarrow$ 2\}\\
			5&\{3$\rightarrow$ 2$\rightarrow$ 1\}\\
			6&\{3$\rightarrow$ 1$\rightarrow$ 2\}\\
			\hline
		\end{tabular}
		\label{table2}
	\end{center}
\end{table}

\begin{table}
	\begin{center}
		\caption{3-paths for sample topology depicted in Fig. \ref{fig:fig2b}.}
		\begin{tabular}{|c|c|c|c|}
			\hline
			\textbf{Label}& \textbf{Path}&\textbf{Label}& \textbf{Path} \\
			\hline
			1&\{1$\rightarrow$ 2\}&7&\{1$\rightarrow$3$\rightarrow$2\}\\
			2&\{2$\rightarrow$ 1\}&8&\{1$\rightarrow$3$\rightarrow$4\}\\
			3&\{3$\rightarrow$ 2\}&9&\{3$\rightarrow$4$\rightarrow$1\}\\
			4&\{3$\rightarrow$ 4\}&10&\{4$\rightarrow$1$\rightarrow$3\}\\
			5&\{4$\rightarrow$ 1\}&11&\{4$\rightarrow$1$\rightarrow$2\}\\
			6&\{4$\rightarrow$ 3\}&12&\{4$\rightarrow$3$\rightarrow$2\}\\
			\hline
		\end{tabular}
		\label{table3}
	\end{center}
\end{table}

\begin{algorithm}
	\caption{\textit{Precompute} x-paths}
	\label{PCXP}
	\begin{algorithmic}[1]
		\break\INPUT {$G=<V,E>$, $x>=1$}
		\OUTPUT {$R=<r>$, $r$ is a set of paths }
		\State{$R=\{\}$;}
		\For{each vertex $v$ in $V$}		
		\For{each vertex $v'$ in $V$}
		\If{$v'$ is a neighbor of $v$}
		\State{$R=R+\{<v,v'>\}$;}
		\EndIf
		\EndFor{}
		\For{each $r$ in $R$}
		\If{($r$ is a neighbor of $v$) \& ($Length(r,v)<=x$)}
		\State{$R=R+\{<r,v>\}$;}
		\EndIf
		\EndFor{}
		\State{$V=V-\{v\}$;}
		\EndFor{}
	\end{algorithmic}
\end{algorithm}

{Alg. \ref{PCXP} provides the process of pre-computing the x-paths. Lines 2-7 of the algorithm find all x-paths with length one. To this end, for each switch $v$ links that that directly connect a switch to $v$ are added to the set of results $R$. In the next step, all x-paths with the length of two are added to the results set $R$ (lines 8-12). To this end, considering each path $r$ in the results set $R$, switches $v$ that have a direct connection to one of the switches in $r$ are added to the set of results $<r,v>$. Thereafter, all 3-paths are considered and so on. At the end, all x-paths with the length of $x$ will be produced.}

\subsection{Chromosomes structure}\label{sec:chrom}
\begin{figure}
	\begin{center}
		\includegraphics[width=0.8\columnwidth]{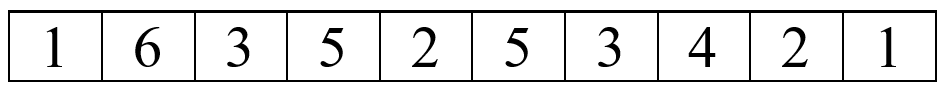}
		\caption{A sample Chromosome for topology depicted in Fig. \ref{fig:fig2a} and 3-paths labeling of Table \ref{table2}.}
		\label{fig:fig3}
	\end{center}
\end{figure}

{In the context of the genetic algorithm each solution is called a \textit{chromosome}. In the proposed algorithm each chromosome (or simply each solution) is an array of $p$ labels (where $p$ is the number of flows in the network), e.g., if there are 10 flows in the topology illustrated in Fig. \ref{fig:fig2a}, then a sample chromosome for 3-path is depicted in Fig. \ref{fig:fig3}. The first element of this chromosome is 1 which means that the selected path for the first flow is \{1$\rightarrow$2\} (based on Table \ref{table2}), therefore, the source of the first flow is switch 1 and the destination is switch 2. Similarly, the sixth element is 5 which means that the selected path for the sixth flow is \{3$\rightarrow$2$\rightarrow$1\}. It should be mentioned that all elements of Table \ref{table2} are calculated based on the topology depicted in Fig. \ref{fig:fig2a}.}

\subsection{Selection structure}\label{sec:SelectionStructure}
After the initial population is created, each chromosome is ranked based on the constraints violations (named as the \textit{fitness function (FF)}). In this way, the amount of traffic that violates the QoS constraints is measured and a proper penalty is assigned to each flow. Due to the complex nature of resource reallocation problems, outstanding individuals may introduce a bias in the early stage of the algorithm. As a result, the algorithm may get on a local optimum. To solve this issue, the parents selection exploit a roulette wheel algorithm \cite{31back1996evolutionary}, which is also known as fitness proportionate selection.

In roulette wheel, let $FF_i$ be the fitness of individual $i$ in the population. The selection probability of $i$-th individual is $p_i=\frac{FF_i}{\sum_{j=1}^{N_{L}} FF_i}$, where $N_{L}$ is the number of individuals in the population. This selection algorithm is chosen since it discards none of the individuals in the population and gives a chance to all of them. As an example, consider that there are 5 different chromosomes (solutions) which have fitness values of \{6.82, 1.11, 8.48, 2.57, 3.08\}, respectively. Therefore, the proportional probability of selecting each chromosome is \{31\%, 5\%, 38\%, 12\%, 14\%\}, respectively. 
 In a nutshell, the selection process consists of two elements: i) fitness function (reported in Alg.~\ref{FF}); and, ii) roulette wheel selection (detailed in Alg.~\ref{RW}). 

In Alg.~\ref{FF}, lines 1-5 compute the imposed traffic to the network by the flow. Line 6 considers the summation of the amount of congestion in all links as the fitness function of the corresponding flow that is returned by the Alg.~\ref{FF}. 
In Alg.~\ref{RW}, lines 3 and 4 calculate the summation of all chromosome fitness values. The cumulative probability of each chromosome is computed in lines 7-10. The loop in line 11 repeats the following instruction until the number of the selected chromosome reaches a predefined population size. In lines 12-20, a random number between $(0,1]$ is generated. In particular, a chromosome with the minimum cumulative probability greater than this random number is selected. 

\begin{algorithm}[H]
	\caption{\textit{FF: Fitness Function}}
	\label{FF}
	\begin{algorithmic}[1]
		\break\INPUT {$CH=<XP>$, $d$, $B=<E,b>$. $CH$ is a set of vertexes (a chromosome), $XP$ is a x-path label, $d$ is the demand size, $B$ is the graph of bandwidth, $E$ is set of edges, and $b$ is the link's capacity}
		\OUTPUT {$F$. fitness associated with the input chromosome}
		\For{each path $XP$ in $CH$}
		\For{each edge $e$ in $XP$}
		\State{reduce the bandwidth of edge $e$ in $B$ with $d$}
		\EndFor
		\EndFor
		\State{$F=sum(b)$;}
		\\
		\Return{$F$}
	\end{algorithmic}
\end{algorithm}

\subsection{Crossover structure}\label{sec:cross}
In genetic algorithms, a crossover is a genetic operator used to vary the features of chromosomes from one generation to the next. The \textit{uniform crossover} uses a fixed mixing ratio between two parents. Unlike one-point and two-point crossover, the uniform crossover enables the parent chromosomes to contribute the gene level rather than the segment level. Therefore, since each chromosome contains several labels, a uniform crossover is exploited in this paper. The corresponding procedure is detailed in Alg.~\ref{UC}. 
\begin{algorithm}
	\caption{\textit{Roulette\_Wheel\_Selection}}
	\label{RW}
	\allowdisplaybreaks
	\begin{algorithmic}[1]
		\break\INPUT {$PO$, which is population}
		\OUTPUT {$S$, which is selected chromosomes}
		\State{$sum=0$;}
		\State{$S=\{\}$;}
		\For{each chromosome $CH$ in $PO$}
		\State{$sum += FF(CH)$;}
		\EndFor
		\State{$sum\_pr=0$;}
		\For{each chromosome $CH_i$ in $PO$}
		\State{$pr_{i} = sum\_pr + (FF(CH_i) / sum)$;}
		\State{$sum\_pr$ += $pr_{i}$;}
		\EndFor
		\While{$Length(S)\leq Length(PO)$}
		\State{$j=1$;}
		\For{$j<=2$}
		\State{$rn = Random(0,1)$;}
		\For{each chromosome $CH_i$ in $PO$}
		\If{$rn$ $>$ $pr_{i}$ \&\& $rn$ $<$ $pr_{i+1}$}
		\State{$S=S+\{CH_i\}$;}
		\EndIf
		\EndFor
		\EndFor
		\EndWhile\\
		\Return{$S$}
	\end{algorithmic}
\end{algorithm}
More in detail, a random number between $(0,1]$ is generated for each gene. If the random number is less than a threshold (i.e., 0.5 in~\cite{2su2016optimization}), then the first child $CH'_1$ receives the corresponding gene from the first parent $CH_1$ and the second child $CH'_2$ receives the corresponding gene from the second parent $CH_2$. Otherwise, the first child receives the gene from the second parent while the second child receives the gene from the first parent.
\begin{algorithm}
	\caption{\textit{Uniform\_Crossover}}
	\label{UC}
	\begin{algorithmic}[1]
		\break\INPUT {$CH_1$, $CH_2$, which are parents (two chromosomes)}
		\OUTPUT {$CH'_1$, $CH'_2$, which are children (two chromosomes)}
		\For{each path in $CH'_1$ and $CH'_2$}
		\If{$Random(0,1)<0.5$}
		\State{$CH'_1$ take the path from $CH_1$}
		\State{$CH'_2$ take the path from $CH_2$}
		\Else{}
		\State{$CH'_1$ take the path from $CH_2$}
		\State{$CH'_2$ take the path from $CH_1$}
		\EndIf
		\EndFor\\
		\Return{$CH'_1$, $CH'_2$}		
	\end{algorithmic}
\end{algorithm}

\subsection{Mutation structure}\label{sec:mut}
\begin{figure}[!htbp]
	\begin{center}
		\includegraphics[width=\columnwidth]{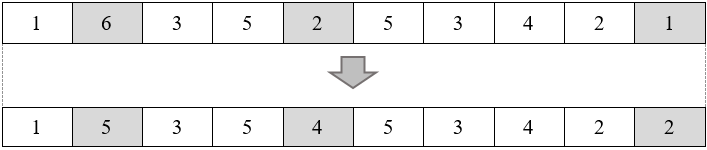}
		\caption{The multipoint mutation.}
		\label{fig:fig6}
	\end{center}
\end{figure}

Mutation is a genetic operator used to maintain genetic diversity from one generation of a population to another one. Due to the fact that the probability of falling into a local optimum in resource reallocation problems are high, CECT uses multipoint mutation operator in which several labels of each chromosome are selected and changed randomly as shown in Fig. \ref{fig:fig6} (i.e., three labels are mutated).

Alg.~\ref{MM} gives an outline of the mutation process. For each gene (which is an x-path label for the corresponding flow), a random number between $(0,1]$ is generated. If the random number is more than the mutation rate $MR$ (line 3), then the corresponding gene is replaced by a feasible gene (line 6), otherwise it may remain unchanged. Feasible genes are sets of x-paths that have similar source and destination with that gene, e.g., if the source and destination of the $i$-th gene that is going to be mutated is $s_i$ and $d_i$, then set of x-paths that have $s_i$ and $d_i$ as their sources and destinations are feasible for this gene.
\begin{algorithm}
	\caption{\textit{Multipoint\_Mutation}}
	\label{MM}
	\allowdisplaybreaks
	\begin{algorithmic}[1]
		\break
		\INPUT {$CH$, $MR$, where $MR$ is mutation rate}
		\OUTPUT {$CH'$, which is the mutated chromosome}
		\State{$H'=\{\}$}
		\For{each gene $r$ in $CH$}
		\If{$Random(0,1)>=MR$}
		\State{$CH'=CH'+\{r\}$}
		\Else
		\State{$CH'$=$CH'$+$Random\{$Feasible Path for this Flow$\}$}
		\EndIf
		\EndFor\\
		\Return{$CH'$}
	\end{algorithmic}
\end{algorithm}
\subsection{Flow table compression structure}\label{sez:4.6}
In this subsection, in order to reduce the computational complexity of CECT, a technique to reduce the size of the flow table is proposed. For the sake of simplicity, all flows belong to the same pair of source and destination called \textit{SF flows}. There is a large number of small flows (i.e., flows with a size less than 10 $Kb/s$) \cite{25benson2012new} in data center networks which can be merged to reduce the computational complexity of the solution. To this end, all SF flows that are smaller than a predefined lower bound are merged. Note that the outcome must be smaller than an upper bound threshold, otherwise it breaks into two or more flows. Interestingly, the increase of the lower bound (fine-grained granularity) reduces the computational complexity while it increases the optimality gap. However, as we will show in the performance evaluation section, the proposed compression method is able to dramatically reduce the number of active flows.

\section{Complexity Analysis}\label{sec:complexityAnalysis}
{In this section the computational and space complexity of the proposed meta-heuristic algorithm is analyzed.}
\subsection{Computational Complexity}
{The algorithm is composed of an online part and an offline part (pre-computation of the x-paths). Since the offline part is related to the network topology and is calculated when the network is configured one time for ever, then we ignore this part of the algorithm in our analyze. CECT is composed of four main subroutines: ranking, selection, crossover, and mutation. Consider $N_F$ as the number of flows, $itr$ as the predefined maximum iteration, $m$ as the maximum length of a path, $L$ as the number of links, $N_p$ as the number of chromosomes, and $c$ as the number of pre-computed paths. The computational complexity of ranking and selection parts are $O(N_F\times(m^2\times \log{c+L}))$ and $O(N_F)$, respectively \cite{26fast}. Both the crossover and mutation steps have a computational complexity of the order $N_F\times m$.
\begin{figure*}[t]
	\begin{center}
		\includegraphics[width=\textwidth]{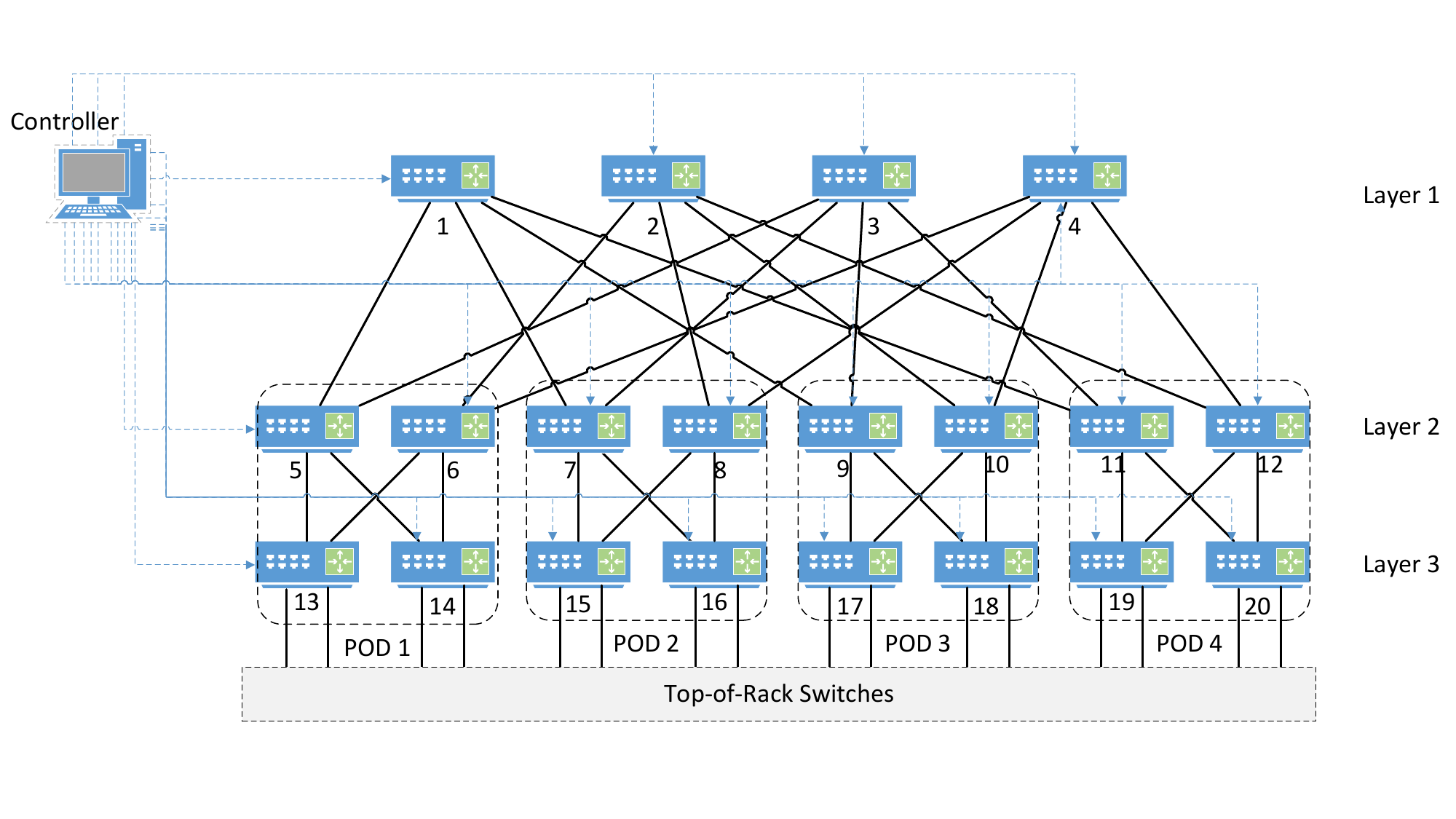}
		\caption{The Fat-tree topology with $k=4$.}
		\label{fig:fig7}
	\end{center}
\end{figure*}
These subroutines are invoked for each pair of chromosomes, consequently the complexity should multiply by $N_p$. Since these parts are executed until a valid solution is achieved or a predefined threshold $itr$ is met, in the worst case the computational complexity of CECT is $O(N_p\times N_F\times itr\times(2m+L+m\times L^2\times \log {c}))$. In our case, the value of $m$, $c$, and $itr$ is selected as 10, 50, and 100, respectively. As a result, computational complexity of CECT is:  
\begin{equation}\nonumber
O(N_p\times N_F\times itr\times m\times L^2\times \log {c}).
\end{equation}
The authors of \cite{abu1999convergence} investigate different approaches of selecting the best population size (number of chromosomes) to be deployed in micro GA algorithms. They propose that the best population size is the square root of the chromosome length. Therefore, considering $N$ as the number of switches, where $N_p=\sqrt{N_F\times \log_2{N}}$. So, CECT computation complexity is:
\begin{equation}\nonumber
O(\sqrt{\log_2{N}}\times {N_F}^{3/2}\times itr\times m\times L^2\times \log {c}).
\end{equation}
The values of $\sqrt{\log_2{N}}$ and $\log {c}$ are less than 10 even for huge networks, therefore, the computational complexity of CECT is as follows:
\begin{equation}\label{eq:eq9}
O(CECT)\triangleq O({N_F}^{3/2}\times itr\times L^2\times m).
\end{equation}
It should be mentioned that $m$ and $itr$ are small values (usually less than 10 and 100, respectively).
}
\subsection{Space Complexity}
{Consider $N_p$ as the number of chromosomes, $IL$ and $CL$ as the length of an integer variable and a character variable, $N_F$ as the number of flows, $m$ as the maximum length of a path, and $c$ as the number of pre-computed paths. We analyze the space complexity of the offline and the online parts separately. In the offline part, CECT should save the table of x-paths and corresponding labels. Each x-path in the worst case consists of $m$ characters where each character contains a switch name. On the other hand, for each x-path there is a corresponding label which is an integer variable. There are $c$ x-paths in the network, therefore, the space complexity of the offline part of the algorithm is $O(c\times (IL+m\times CL))$.} 

{On the other hand, in the online part of the algorithm, there are $N_p$ chromosomes, each chromosome consists of $N_F$ integer variables (labels). Therefore, the space complexity of the online part of the algorithm is $O(N_F\times N_p\times IL)$.}

\section{Simulated scenarios and considered performance metrics}\label{sez:6}
In this section, we describe the considered test scenarios and the adopted performance metrics.

\subsection{Setup Description}\label{sez:5.1}
The proposed analytical model is evaluated on the network topology shown in Fig. \ref{fig:fig7} in order to simulate the SCDC. The illustrated topology called fat-tree topology is a scalable data center network architecture that is universally adopted \cite{27leiserson1985fat}. All simulations are carried out on a desktop equipped with Intel Core 2 dual 2.6 $GHZ$ CPU and 4.0 $GB$ RAM.  In our simulations, we consider 32 port switches (e.g., NEC PF5340-32QP) on the 3$^{rd}$ layer of the topology (switches 13-20) and 48 port top-of-rack switches (e.g., Cisco Catalyst 4948 Switch), this topology can support up to 11.28 $K$ servers. We exploit a network traffic pattern which can be found in \cite{28benson2010network}. Due to lack of information about the IP layout, we assign hosts to the switches randomly. In addition, the access and aggregation switches are classified as POD I. The probability of leaving the originating POD for each flow is considered as PLR parameter, e.g., $PLR=0$ means that all flows stay in their originating POD. We use Mininet \cite{29mininet} along with POX controller \cite{30pox} to emulate the network. In the sequel, network throughput of ECMP \cite{23-5454063} and CECT are discussed to show the impact of the proposed scheme on the network performance. {ECMP is selected as a comparisons with our solution (i.e., CECT) for two reasons: i) ECMP and CECT as multi-path routing method distribute packets across multiple links in the network in such away to preserve the load balancing, and, ii) ECMP is considered as an interesting and prevalent real method which is applied in large data centers and it is implemented as a common routing protocol in Mininet \cite{29mininet}.}

\subsection{Performance Metrics}\label{sez:5.2}
In the carried out simulations, the following three performance metrics have been numerically evaluated:
\begin{enumerate}
	\item[(i)] \emph{Throughput}: it is the rate of successful message delivery over a communication channel in the SCDC; 
	\item[(ii)] \emph{Data Transfer}: it is the average amount of the data transferred through a link in the SCDC;
	\item[(iii)] \emph{Packet Loss}: it is the network congestion metric that is the percentage of packets lost with respect to packets sent in the SCDC.
\end{enumerate}

It should be mentioned that Wireshark is used to capture the traffic in all of the hosts and switches. Thereafter, all captured traffics are merged to calculate the mentioned parameters.

\section{Performance Evaluation and Comparisons}\label{sez:7}
In this section, we test and compare the performance of the proposed CECT algorithm against the corresponding one, namely, the ECMP \cite{23-5454063} algorithm.

\subsection{Throughput}\label{sez:6.1}
\begin{figure}[ht]
	\begin{center}
		\includegraphics[width=\columnwidth]{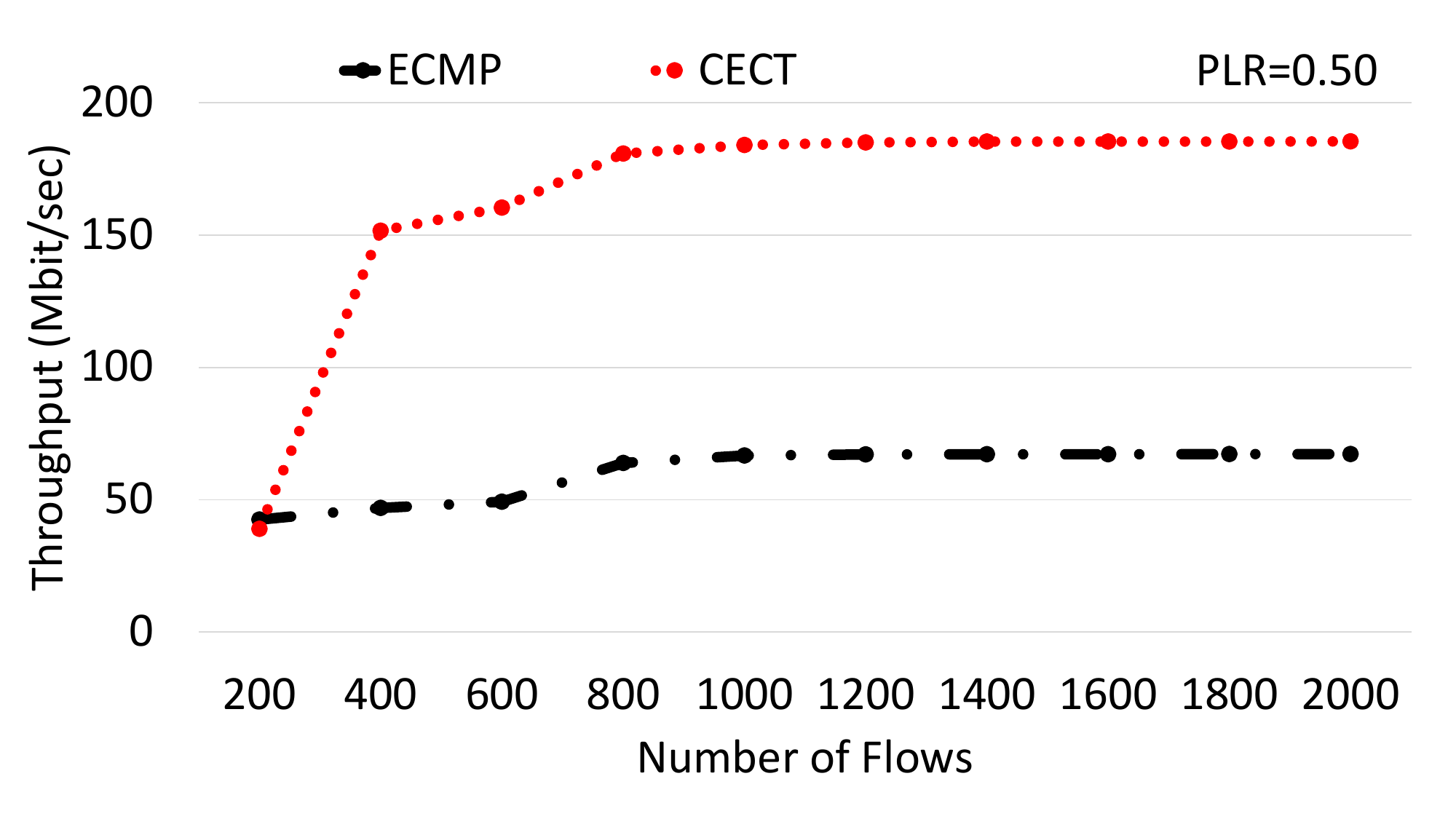}
		\caption{Total network throughput.}
		\label{fig:fig8}
	\end{center}
\end{figure}

The first group of carried out tests aims to evaluate and compare the throughput of CECT and ECMP (Equal Cost Multi-Path) \cite{23-5454063} routing algorithms. Hence, in our emulation, ECMP considers the path length as the cost of that path which means that it uses the shortest-path algorithm to find paths and distributes the traffic between these paths. The obtained numerical results for the total number of flows are reported in Fig. \ref{fig:fig8}. 

In Fig. \ref{fig:fig8}, we increase the total number of flows in the network from 200 flows up to 2000 flows by adding 200 new flows in each iteration and calculate the network throughput. To this end, we run \textit{TCP dump} on all of the hosts in the network and merge these TCP-dumps to calculate the total network throughput. As can be seen, since growing the number of flows increases the network traffic load, the probability of congestion in the network is higher. Therefore, the superiority of CECT is more evident in a large number of network flows. Based on the emulation results, CECT improves the network throughput up to 3x compared to the ECMP.

\begin{figure*}
	\centering
	\begin{subfigure}[t]{0.65\columnwidth}
		\centering
		\includegraphics[width=\columnwidth]{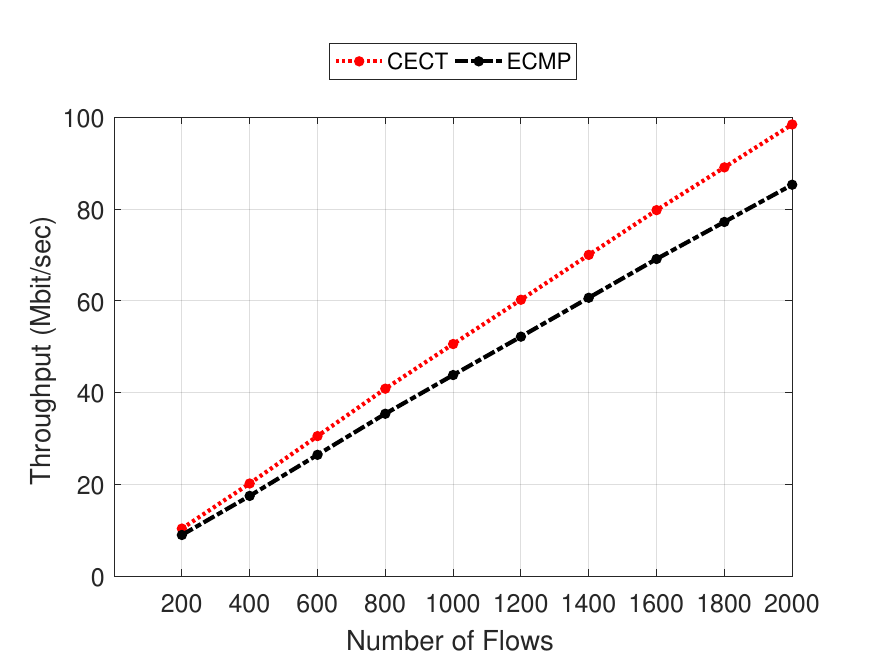}
		\caption{Micro Flows.}
		\label{fig:microFlows}
	\end{subfigure}
	\hfill
	\begin{subfigure}[t]{0.65\columnwidth}
		\centering
		\includegraphics[width=\columnwidth]{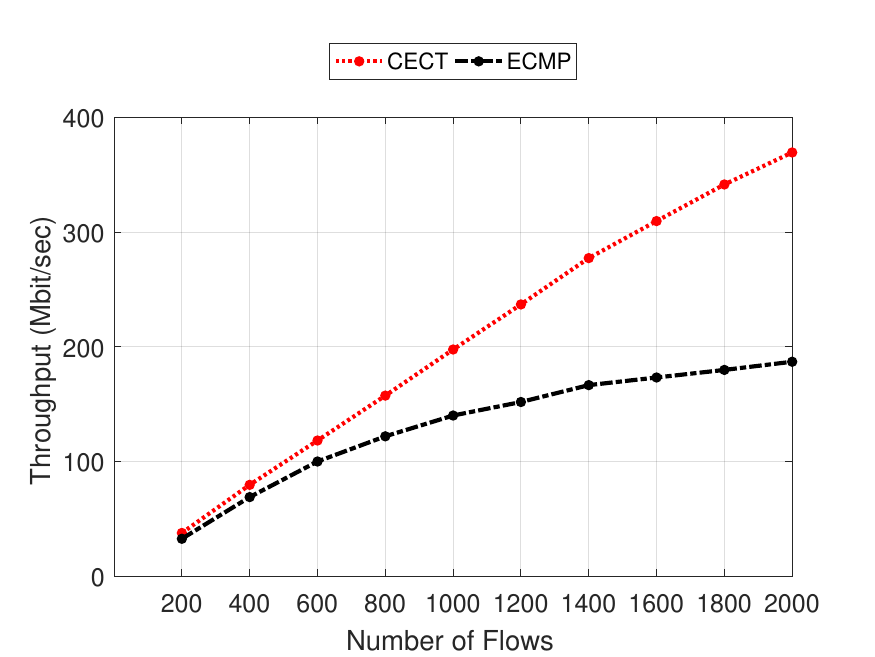}
		\caption{Small Flows.}
		\label{fig:smallFlows}
	\end{subfigure}
	\hfill
	\begin{subfigure}[t]{0.65\columnwidth}
		\centering
		\includegraphics[width=\columnwidth]{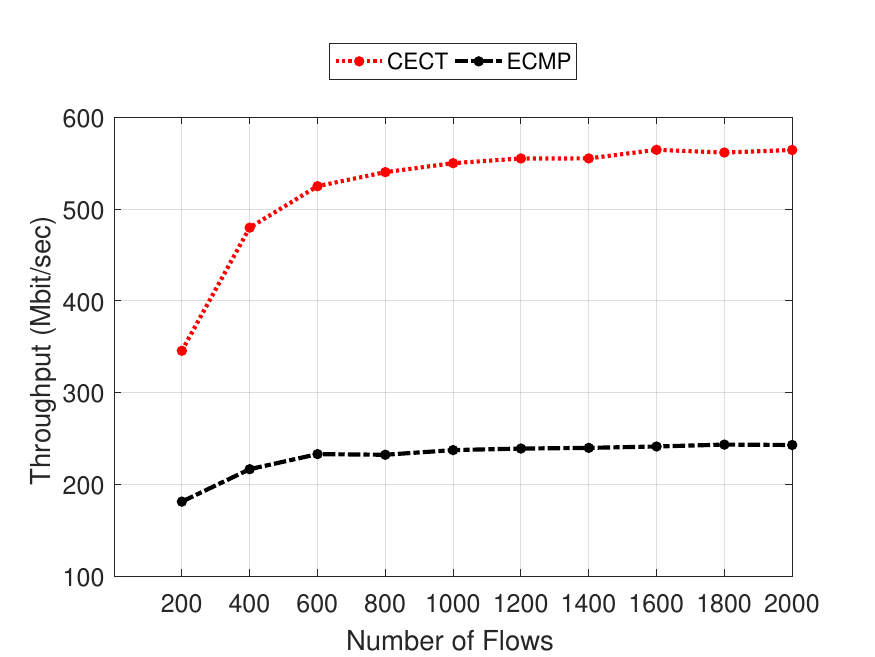}
		\caption{Medium Flows.}
		\label{fig:mediumFlows}
	\end{subfigure}
	\hfill
	\begin{subfigure}[t]{0.65\columnwidth}
		\centering
		\includegraphics[width=\columnwidth]{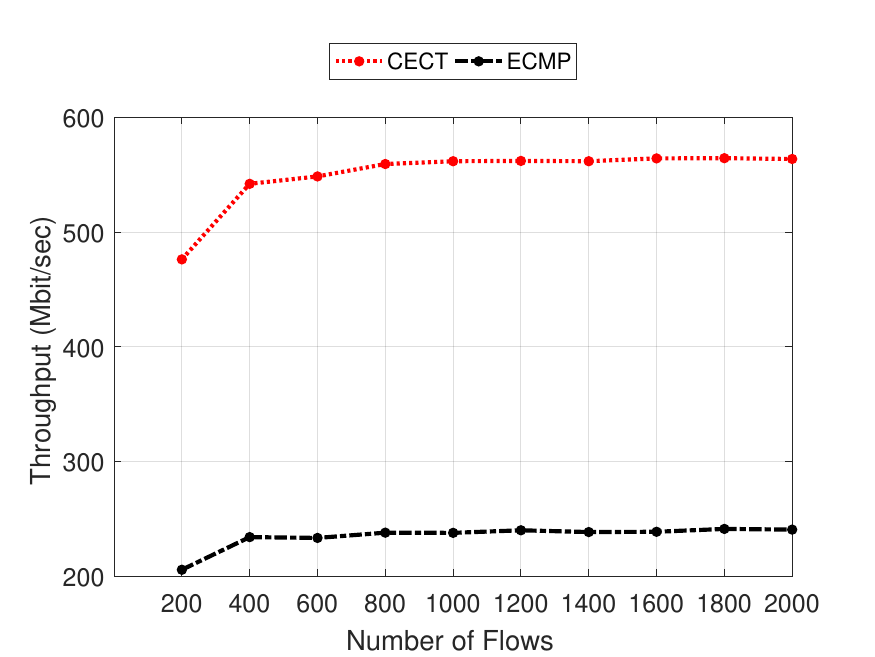}
		\caption{Big Flows.}
		\label{fig:bigFlows}
	\end{subfigure}
	\hfill
	\begin{subfigure}[t]{0.65\columnwidth}
		\centering
		\includegraphics[width=\columnwidth]{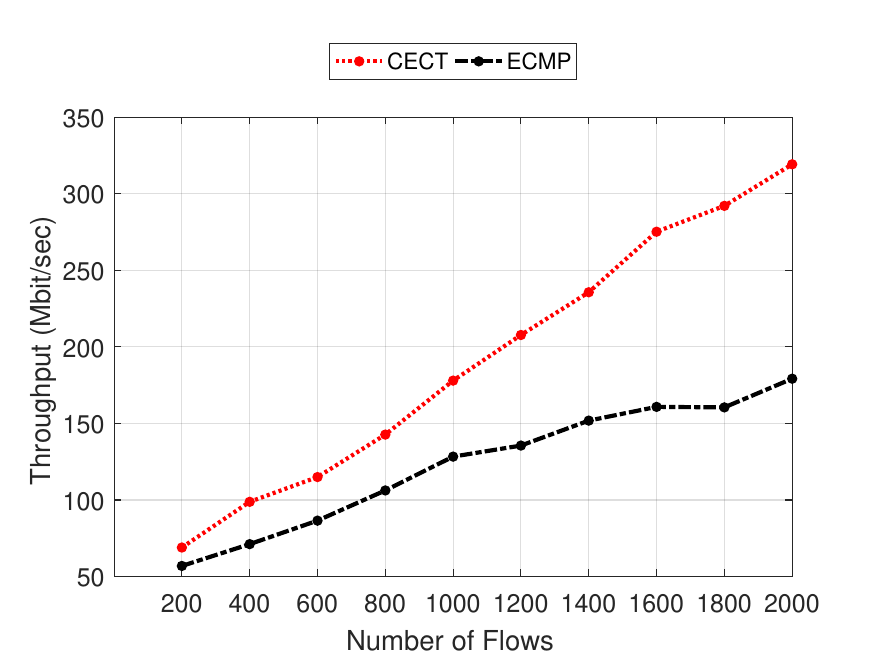}
		\caption{Real Network Traffic.}
		\label{fig:thrg645}
	\end{subfigure}
	\caption{Network Throughput, fat-tree k=6, 45 switches.}
	\label{fig:ThroughputScalability}
\end{figure*}

{In the following, in order to investigate the impact of the network size and traffic pattern over the proposed scheme, we evaluate the network throughput over a network topology with 45 switches (i.e., fat-tree k=6). Additionally, we implement a network packet generator to generate different traffic patterns. To this end, we use MATLAB programming language to generate the traffic pattern with different flow sizes and simulate the network environment. Furthermore, we use a real network traffic with 4 different traffic patterns where the average rate of flows are micro (Fig.~\ref{fig:microFlows}), small (Fig.~\ref{fig:smallFlows}), medium (Fig.~\ref{fig:mediumFlows}), and big (Fig.~\ref{fig:bigFlows}). Consider $b$ as the ratio of the flow rate to link bandwidth. We consider flow $f$ a micro flow, if $b_f=0.005$. Similarly, small, medium, and big flows are flows with $b_f$ equal to 0.02, 0.2, and 0.5, respectively. As can be seen, the superiority of CECT over ECMP increases whether the size or the number of traffic flows increases.
}

\subsection{Data transfer}\label{sez:6.2}
\begin{figure}
	\centering
	\begin{subfigure}[t]{\columnwidth}
		\centering
		\includegraphics[width=\columnwidth]{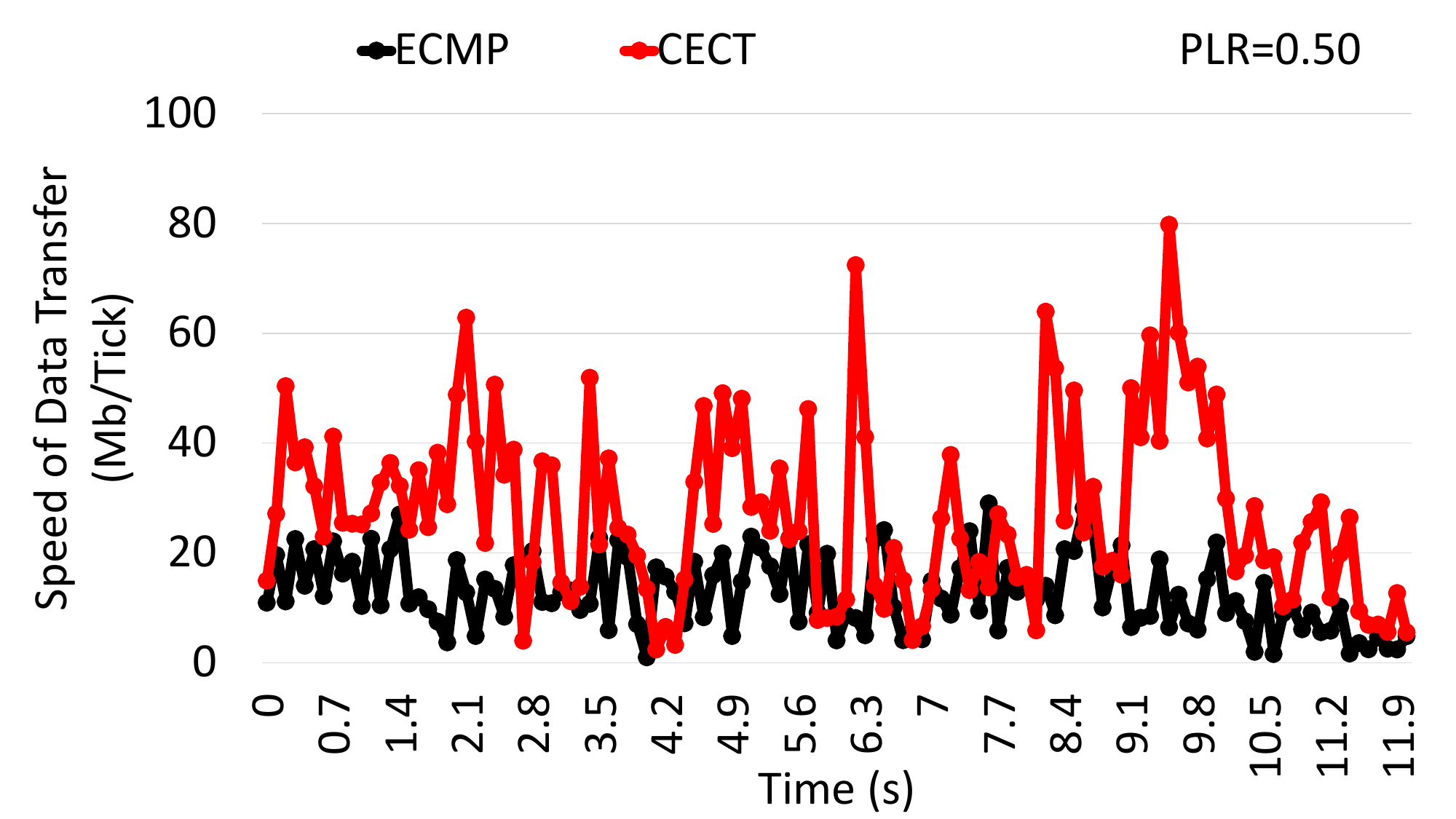}
		\caption{Traffic transferred.vs. time interval.}
		\label{fig:fig9a}
	\end{subfigure}
	\hfill
	\begin{subfigure}[t]{\columnwidth}
		\centering
		\includegraphics[width=\columnwidth]{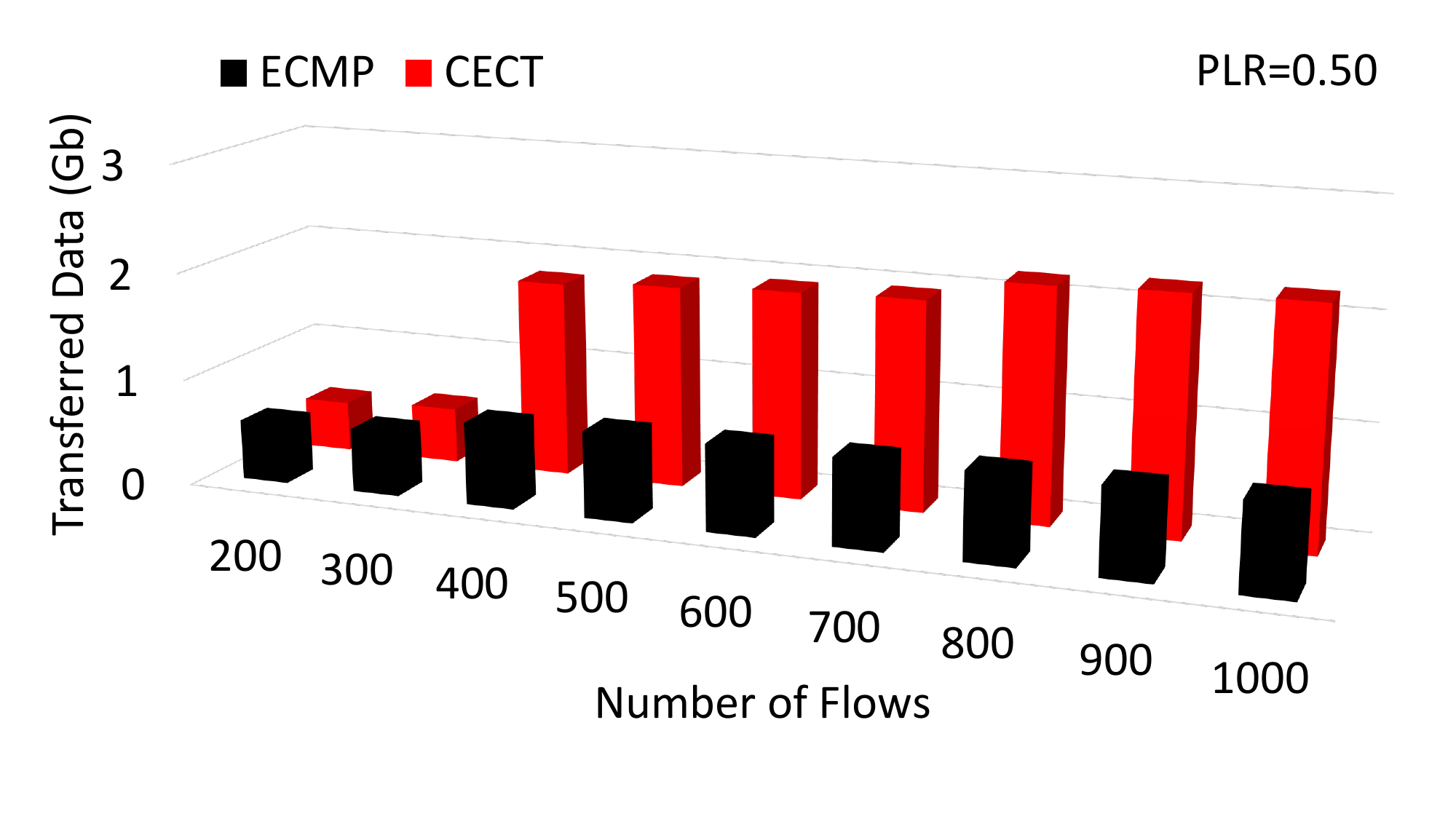}
		\caption{Traffic transferred.vs. Number of flows.}
		\label{fig:fig9b}
	\end{subfigure}
	\caption{Network Traffic.}
	\label{fig:fig9}
\end{figure}

The second group of numerical tests focuses on the communicated traffic in the network in each time interval and the transferred data versus the number of flows between CECT, ECMP routing strategies that are presented in Fig. \ref{fig:fig9}. Hence, Fig. \ref{fig:fig9a} shows the number of bytes that are communicating in each time interval. Based on this figure, the superiority of CECT over ECMP (the shortest path algorithm is used as the cost function) is clear. Due to greedy nature of TCP connections, each flow tries to obtain as much bandwidth as possible, therefore, the TCP connections try to reach the maximum speed of data transfer. However, since increasing the traffic rate increases the probability of congestion, the traffic rate rises and then drops periodically. On the other hand, Fig. \ref{fig:fig9a} presents the transferred data versus the number of flows in the network. As can be seen, increasing the number of flows increases the gap between the results of ECMP and CECT. This happens because increasing the number of flows, increases the probability of congestion in the network and makes the impact of the rerouting algorithm clearer.

It should be mentioned that in our emulation, each flow tries to communicate a special amount of traffic (e.g., flow 1: 10 $Mb$, flow 2: 400 $Kb$, flow 3: 100 $Mb$, etc.). 

\subsection{Packet Loss}\label{sez:6.3}
\begin{figure}
	\begin{center}
		\includegraphics[width=\columnwidth]{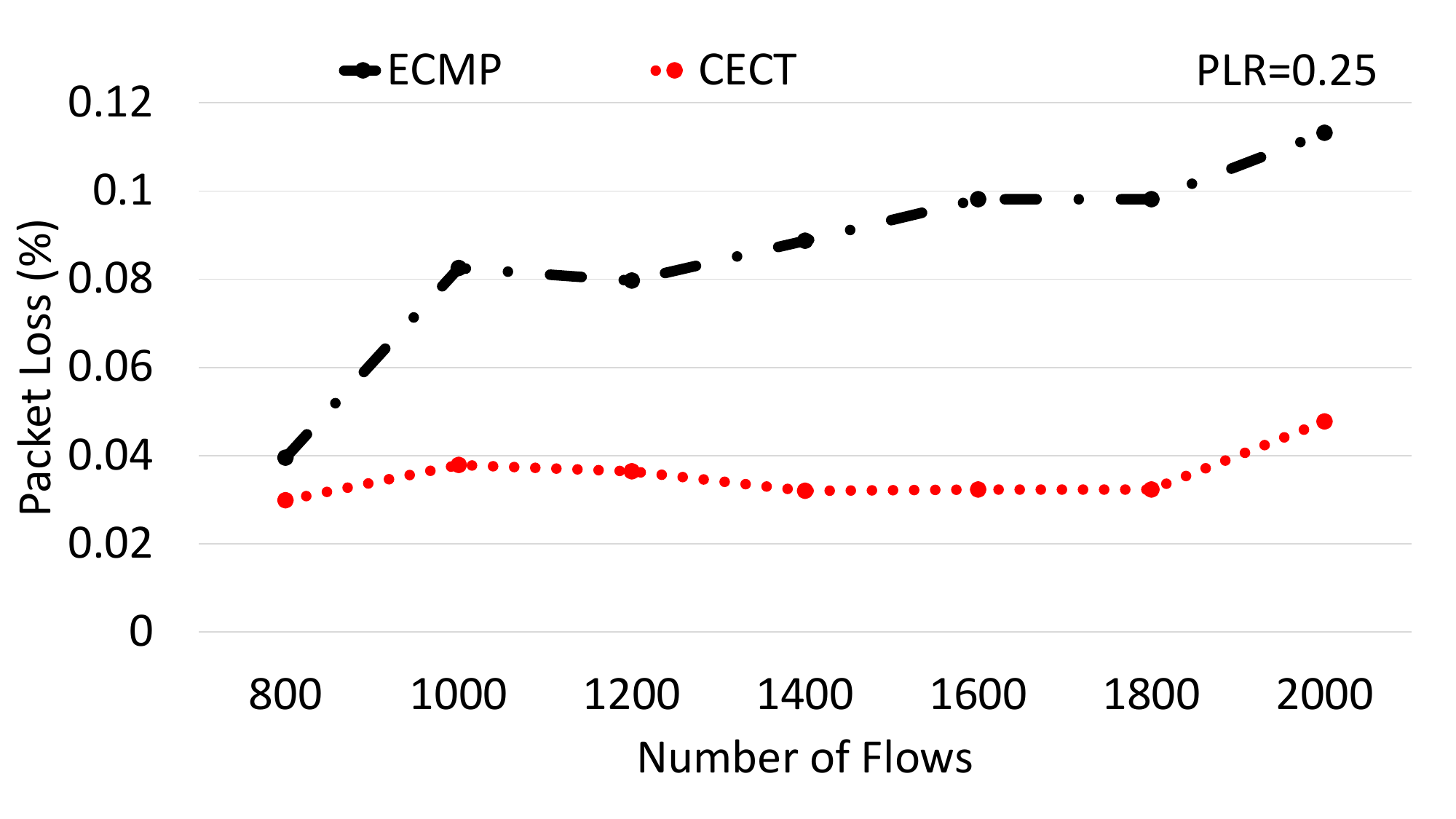}
		\caption{Packet loss.}
		\label{fig:fig10}
	\end{center}
\end{figure}

The third group of simulations aims at evaluating the packet loss of the network. The obtained numerical results (expressed in terms of the multiple number of flows) are reported in Fig. \ref{fig:fig10}. The percentage of packet loss versus the number of flows in presented in the mentioned figure. It is clear from the figure that increasing the number of flows increases the packet loss in both approaches, however, the percentage of packet loss in the proposed algorithm is sufficiently lower in compared with the packet loss of ECMP. As a result, CECT decreases the packet loss up to 2x compared to traditional approach (i.e., ECMP).

\subsection{Execution Time}
\begin{figure}
	\centering
	\begin{subfigure}[t]{0.9\columnwidth}
		\centering
		\includegraphics[width=\columnwidth]{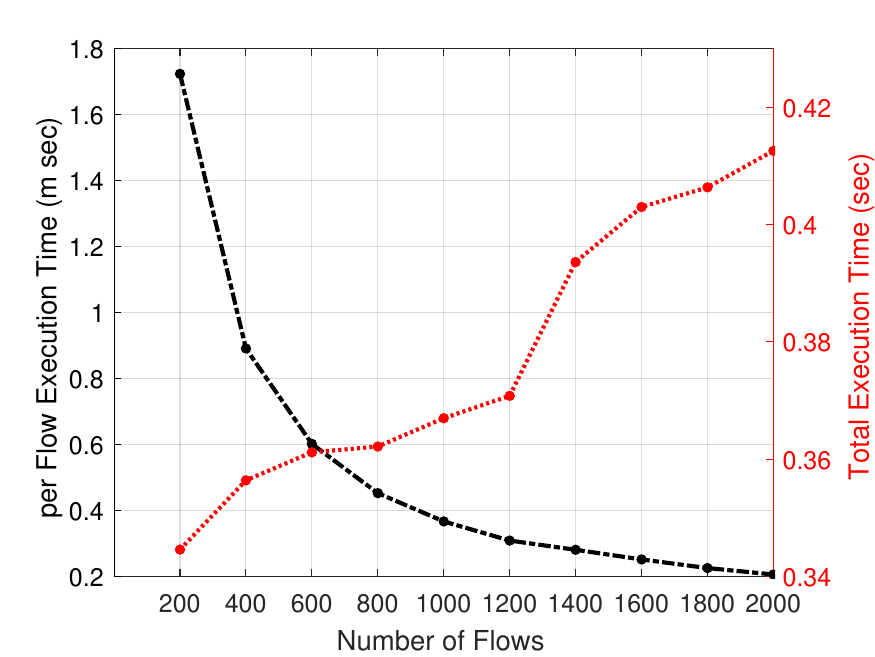}
		\caption{Fat-tree K=4, 20 switches.}
		\label{fig:cmplxK4}
	\end{subfigure}
	\hfill
	\begin{subfigure}[t]{0.9\columnwidth}
		\centering
		\includegraphics[width=\columnwidth]{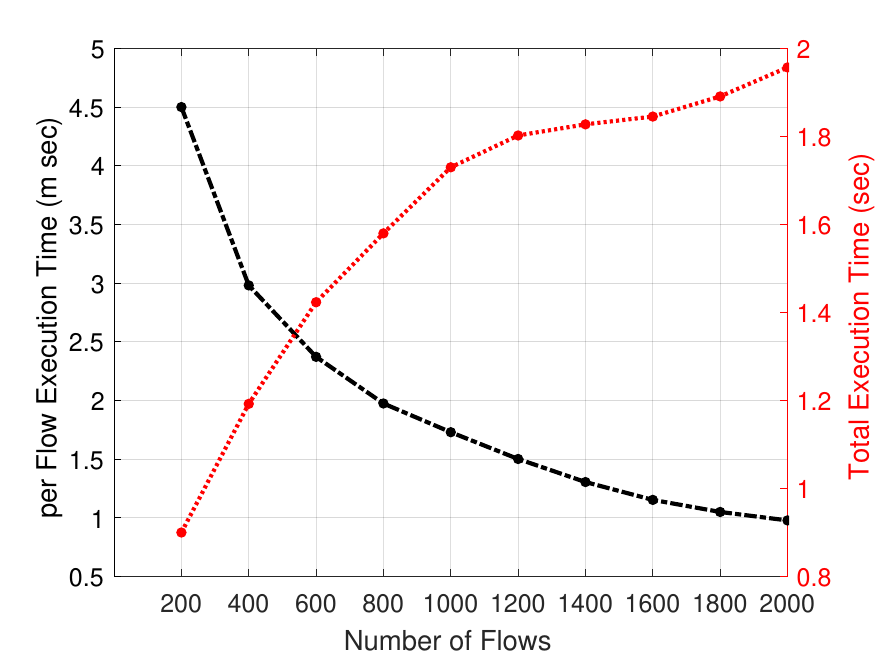}
		\caption{Fat-tree K=6, 45 switches.}
		\label{fig:cmplxK6}
	\end{subfigure}
	\caption{Execution time of CECT.}
	\label{fig:time}
\end{figure}

{In order to investigate the impact of network size and flows number on the execution time of CECT, we exploit a network topology with 45 switches and 2000 flows to test CECT and the results are depicted in Fig.~\ref{fig:time}. In the right side of the figure, \textit{total execution time} of CECT versus the number of flows is presented. Correspondingly, on the left side, \textit{per flow execution time} versus the number of flows is illustrated. As can be seen, the total execution time for 20 and 45 switches are less than 0.42 and 2 second and the execution time per each flow is less than 1.8 and 4.5 milli-second.}

{Since the execution times is completely dependant on the programming language, the configuration of the PC which hosts the network controller, and the optimality of the implementation, therefore, we mathematically analyzed the computation and space complexity of CECT in Section \ref{sec:complexityAnalysis}.}

\section{Conclusions and Future Work}\label{sez:8}
In this work, an efficient resource reallocating algorithm for software-defined data centers was introduced. In this way, the problem was mathematically formulated and an optimal scheme was proposed to solve the corresponding optimization problem. Since the computational complexity of the proposed solution is high, we proposed a meta-heuristic approach based on genetic algorithm, called CECT, to propose a sub-optimal solution which has a low computational complexity. The computational complexity of CECT was discussed and showed that it is applicable for real-world networks. Additionally, CECT was compared with ECMP from throughput, data transfer, and packet loss perspective. Emulation results show that CECT improves the total network throughput up to 3x while the packet loss is decreased up to 2x. 
Future work would be dedicated to minimizing the side effect of network reconfiguration. Additionally, one can minimize the network energy consumption by chaining the objective function. In this way, the mathematical formulation should be extended in a way that some new constraint implement the network energy consumption model.


\bibliographystyle{spmpsci}
\bibliography{CECTbib}

\end{document}